\documentstyle[11pt,epsf,float,psfig,paspconf]{article}

\index{Source!SgrA*}
\index{accretion}
\index{black hole physics}

\begin{document}
\def\eps@scaling{.95}
\def\epsscale#1{\gdef\eps@scaling{#1}}
\def\plotone#1{\centering \leavevmode
\epsfxsize=\eps@scaling\columnwidth \epsfbox{#1}}

\title{3D Simulation of the Gas Dynamics in the Central Parsec of the Galaxy}
\author{R.F. Coker\altaffilmark{1}} 
\affil{Physics Department, The University of Arizona, Tucson, AZ  85721, USA}

\author{Fulvio Melia\altaffilmark{2}}
\affil{Physics Department and Steward Observatory, The University 
of Arizona, Tucson, AZ  85721, USA}

\altaffiltext{1}{NASA GSRP Fellow.}
\altaffiltext{2}{Presidential Young Investigator.}

\begin{abstract}
It is thought that many characteristics of the gaseous features within
the central parsec of our Galaxy,
are associated with the accretion of 
ambient plasma by a central concentration of mass.  Using a 3D hydrodynamical
code, we have been simulating this process in order to realistically model
the gaseous flows in the center of our Galaxy.  In the most recent simulation,
we have taken into account the multi-point-like distribution of stellar wind
sources, as well as the magnetic heating and radiative cooling of these stellar
winds.  As expected, we find that the structure of the flow is significantly
different from that due to a uniform medium.  We also investigate the possibility
that Sgr A* is due to a distributed mass concentration instead of the canonical
point mass of a black hole.  We discuss the physical state of the accreting
gas and how our results suggest that Sgr A* is unlikely to be associated with
a ``dark cluster''.
\end{abstract}


\section{Introduction}
The hydrodynamics of the interstellar medium (ISM) within the central parsec 
of the Galaxy has long been thought to be dominated by the gravitational potential due
to a central mass concentration (for a review, see Mezger, Duschl,
\& Zylka, 1996).  For example, the Western Arc, one of the dominant kinematic
features of the region, appears to be in circular rotation about Sgr A*,
a unique compact radio source whose lack of proper motion suggests
it lies at the dynamical heart of the Galaxy (Roberts \& Goss, 1993; Backer, 1994).
Stellar motions (Genzel, et. al., 1997), gas kinematics (Herbst, et. al., 1993),
and velocity dispersion measurements (Eckart \& Genzel, 1998) together suggest the
presence of $\sim 2.6\times 10^6 M_\odot$ of dark mass located within 
$\sim .01$ parsecs of Sgr A*.  Note that the Galactic center (GC) is at a 
distance of $\sim 8$ kpc so that $1$ arcsecond $\sim 0.04$ parsecs.

However, showing that the GC must contain a centralized dark
mass concentration does not necessarily imply that it is in the form of a single
compact object nor does it imply that Sgr A* must be associated with it.
Stellar kinematic arguments (Genzel et. al., 1996) rule out a distribution of
neutron stars or white dwarfs.  One possibility is that the dark mass distribution
consists of $\sim 10 M_\odot$ black holes; cluster evolution calculations (Lee, 1995)
have shown that this is at least feasible although stability arguments (Maoz, 1998)
suggest the cluster lifetime would be $\sim 10^8$ years, considerably less than
the age of the Galaxy.  In this paper, we wish to determine if, stability
arguments aside, the spectrum
of Sgr A* could be due to a plasma trapped within the potential well of a
dark cluster of arbitrary objects. 

In addition to large scale gaseous features, there is ample evidence for the
existence of rather strong stellar winds in and around Sgr A* itself.  The key
wind sources appear to be the cluster of mass-losing, blue, luminous stars 
comprising the IRS 16 assemblage, located within several arcseconds of Sgr A*.
A variety of observations over the years (for a review see Morris \& Serabyn, 1996)
provide clear evidence of a hypersonic wind, with a velocity of $v_w \sim 500-1000
$~km s$^{-1}$, a number density $n_w \sim 10^{3-4}$ cm$^{-3}$, and a total mass loss
rate $\dot M_w \sim 3-4\times 10^{-3} M_\odot~$yr$^{-1}$ pervading the inner parsec
of the Galaxy.  If the dark matter is distributed, it is likely that a portion of
this wind is captured by the dark compact cluster and that it settles within the
cluster's potential well.  Although the potential well of a cluster does not include 
a cusp such as that due to a black hole, the trapped plasma might conceivably still
account for at least some of Sgr A*'s radiative characteristics.

In \S 2 we discuss the simulation of the gas flow through a distributed dark
mass cluster.  In \S 3 we describe the resulting spectrum and present
some preliminary semi-analytical spectral calculations based on Sgr A* being a
point mass while in \S 4 we summarize our analysis.

\section{The 3D Hydrodynamical Model}

In the classical Bondi-Hoyle scenario (Bondi \& Hoyle, 1944), the mass accretion
rate for a uniform hypersonic adiabatic gas flowing past a centralized mass is
\begin{equation}\label{mdot}
\dot M_{BH} = \pi {R_A}^2 m_H n_w v_w\;,
\end{equation}
where $R_A \equiv 2 G M / {v_w}^2$ is the accretion radius and $M$ is the mass
of the centralized object(s).  At the GC, for the conditions described in the
Introduction, we would therefore expect an accretion rate $\dot M_{BH} \sim 
10^{21-22}$g sec$^{-1}$, with a capture radius $R_A \sim 0.01-0.02~$pc.  
Since this accretion rate is sub-Eddington
for a $\sim$ one million solar mass concentration, the accreting gas is
unimpeded by the escaping radiation field and is thus essentially in hydrodynamic
free-fall starting at $R_A$.  Our initial, simplistic, numerical simulations of 
this process, where we assume a point object and uniform flow (Ruffert \& Melia, 
1994; Coker \& Melia, 1996)
have verified these expectations.

\subsection{The Stellar Wind Sources}

The GC wind, however, is unlikely to be uniform since the winds from many stars contribute to
the mass ejection.  We assume that the early-type stars
enclosed (in projection) within the Western Arc, the Northern Arm, and the Bar produce
the observed wind.  Thus far, 25 such stars have been identified (Genzel, et. al., 1996),
though the stellar wind characteristics of only 8 have been determined from their
He I line emission (Najarro, et. al., 1997); the relavant characteristics of these 25 stars are
summarized in Table 1.  Two sources, IRS 13E1 and IRS 7W, seem to
dominate the mass outflow with their high wind velocity ($\sim 1000 $~km sec$^{-1}$) and a mass
loss rate of more than $2\times10^{-4}\;M_\odot$ yr$^{-1}$ each.
Unfortunately, the temperature of the stellar winds is not well known, and
so for simplicity we have assumed that all the winds are Mach 30; this corresponds to
a temperature of $10^{4-5}$K.  In addition, for the sources that are used in these calculations,
their location in $z$ (i.e., along the line of sight) is determined randomly with the
condition that the overall distribution in this direction approximately matches that in
$x$ and $y$.  

\begin{table}
\caption{Parameters for Galactic Center Wind Sources}
\begin{center}\scriptsize
\smallskip
\begin{tabular}{lccccc}
Star & x\tablenotemark{a} (arcsec) & y\tablenotemark{a} (arcsec) & 
z\tablenotemark{a} (arcsec) & v (km sec$^{-1}$) & 
$\dot M({10}^{-5}\;M_\odot$ yr$^{-1}$)\\
\tableline
 IRS 16NE  & -2.6 &  0.8 &  5.5 &  550 &  9.5 \\
 IRS 16NW  &  0.2 &  1.0 &  7.3 &  750 &  5.3 \\
 IRS 16C   & -1.0 &  0.2 & -7.1 &  650 & 10.5 \\
 IRS 16SW  & -0.6 & -1.3 &  4.9 &  650 & 15.5 \\
 IRS 13E1  &  3.4 & -1.7 & -1.5 & 1000 & 79.1 \\
 IRS 7W    &  4.1 &  4.8 & -5.1 & 1000 & 20.7 \\
 AF        &  7.3 & -6.7 &  8.5 &  700 &  8.7 \\
 IRS 15SW\tablenotemark{b}  &  1.5 & 10.1 &      &  700 & 16.5 \\
 IRS 15NE\tablenotemark{b}  & -1.6 & 11.4 &      &  750 & 18.0 \\
 IRS 29N\tablenotemark{c}   &  1.6 &  1.4 &  3.5 &  750 & 12.9 \\
 IRS 33E\tablenotemark{c}   &  0.0 & -3.0 &  1.5 &  750 & 12.9 \\
 IRS 34W\tablenotemark{c}   &  3.9 &  1.6 & -6.4 &  750 & 12.9 \\
 IRS 1W\tablenotemark{c}    & -5.3 &  0.3 &  7.8 &  750 & 12.9 \\
 IRS 9NW\tablenotemark{cd}  & -2.5 & -6.2 & -3.8 &  750 & 12.9 \\
 IRS 6W\tablenotemark{c}    &  8.1 &  1.6 &  3.6 &  750 & 12.9 \\
 AF NW\tablenotemark{cd}    &  8.3 & -3.1 & -2.1 &  750 & 12.9 \\
 BLUM\tablenotemark{b}      &  9.2 & -5.0 &      &      &      \\
 IRS 9S\tablenotemark{b}    & -5.5 & -9.2 &      &      &      \\
 Unnamed 1\tablenotemark{b} &  1.3 & -0.6 &      &      &      \\
 IRS 16SE\tablenotemark{b}  & -1.4 & -1.4 &      &      &      \\
 IRS 29NE\tablenotemark{b}  &  1.1 &  1.8 &      &      &      \\
 IRS 7SE\tablenotemark{b}   & -2.7 &  3.0 &      &      &      \\
 Unnamed 2\tablenotemark{b} &  3.8 & -4.2 &      &      &      \\
 IRS 7E\tablenotemark{b}    & -4.2 &  4.9 &      &      &      \\
 AF NWW\tablenotemark{b}    & 10.2 & -2.7 &      &      &      
\end{tabular}
\end{center}
\tablenotetext{a}{Relative to Sgr A* in l-b coordinates (negative x
is east and negative y is south of Sgr A*)}
\tablenotetext{b}{Star not used in these calculations}
\tablenotetext{c}{Wind velocity and mass loss rate fixed (see text)}
\tablenotetext{d}{Star position changed slightly due to finite physical resolution}
\end{table}

The sources are assumed to be stationary over the duration of the simulation.
The stars without any observed He I line emission have been assigned
a wind velocity of $750$ km sec$^{-1}$ and an equal mass loss rate chosen such that the total mass
ejected by the 14 stars used here is equal to $3\times10^{-3}\;M_\odot$
yr$^{-1}$.  Note that although we have matched the overall mass outflow rate
to the observations, we have only used 14 of the 25 stars in the sample.
There are two principal reasons for this: (1) stars further away than
10 arcsec (in projection) from Sgr A* are outside of our volume of
solution and therefore could not be included, and (2) due to our
computational resolution limits, we needed to avoid excessively
large local stellar densities.

In a complex flow, generated by many wind sources, the wind
velocity and density are not uniform, so the accretion radius is not
independent of angle.  To set the length scale for the
simulation, we shall therefore adopt the value $R_A = .018 $pc
(for which 1$^{\prime\prime}$ = 2.3 $R_A$) as a reasonable
mean representation of this quantity.

\subsection{The Dark Cluster Potential}
We wish to study the emission characteristics of a hot, magnetized plasma ``trapped''
within the dark cluster's gravitational potential well.
Following Haller \& Melia (1996), we will represent the gravitational
potential of the dark cluster with an ``$\eta$-model'' (Tremaine, et. al., 1994).
This function represents an isotropic mass distribution with a single
parameter. 
We here restrict our examination to the case $\eta=2.5$ since this
provides the closest approximation to a King model that is physically
realizable (i.e., a nonnegative distribution function).  We scale
the mass so that $2\times10^6\;M_\odot$ are enclosed within $0.01$ pc
and the total integrated mass of the dark cluster
is $2.7\times10^6\;M_\odot$.  Thus, writing $r$ in units of 
$R_A$ we get for the enclosed mass as a function of $r$,
\begin{equation}
M_\eta(r) = 2.7\times10^6 \left({{13.84r}\over{1+13.84r}}
\right)^{5/2} \;M_\odot\;.
\end{equation}
A more recent assessment of the enclosed mass (Genzel, et. al., 1997) places
a yet more rigorous constraint on the possibility of a distributed dark
matter component.  These newer observations may indeed invalidate the idea
that {\sl any} realistic stable distribution of objects can
account for the observed gravitational potential.  

\subsection{The Hydrodynamics Code}
We use a modified version of the numerical finite difference 
algorithm ZEUS, a general
purpose code for MHD fluids developed at NCSA (Stone \& Norman, 1992; Norman, 1994).
The code was run on the massively parallel Cray T3E at NASA's Goddard
Space Flight Center under the High Performance Computing Challenge program.
Zone sizes are
geometrically scaled by a factor of 1.02 so that the central zones are
$\sim 20$ times smaller than the outermost zones, mimicking
the ``multiply nested grids'' arrangement used by other researchers
(e.g., Ruffert \& Melia, 1994).  This allows for maximal resolution of the central
region (within the computer memory limits available) while sufficiently
resolving the wind sources and minimizing zone-to-zone boundary effects.
The total volume is $(40 R_A)^3$ or $\sim(0.7 $pc$)^3$ with the center of
the spherically symmetric dark cluster distribution being located at the origin.

The density of the gas initially filling the volume of solution is set to
a small value and the velocity is set to zero,
while the internal energy density is chosen such that
the initial gas temperature is $\sim10^2$ K.  Free outflow conditions are
imposed on the outermost zones and each time step is determined by the
Courant condition with a Courant number of 0.5.  The 14 stellar wind
sources are modeled by forcing the velocity in 14 subregions of 125 zones
each to be constant with time while the densities in these subvolumes are
set so that the total mass flow into the volume of solution from each source
is given by Table 1.  Also, the magnetic field of the winds
is assumed to always be at equipartition with the thermal energy density.  
The angular momentum and mass accretion
rates are calculated by summing the relevant
quantity in zones located within $0.1 R_A$ of the origin.

We assume the magnetic field is perfectly tangled and thus ignore the effects 
of the magnetic field on the large scale kinematics.
We take the medium to be an adiabatic polytropic gas, with $\gamma = 5/3$.  Building on
previous work (Melia, 1994; Coker \& Melia, 1997), we have included a first order
approximation to magnetic dissipative heating as well as an accurate expression
for the cooling due to magnetic bremsstrahlung, thermal bremsstrahlung, line emission,
radiative recombination, and 2 photon continuum emission for a gas with cosmic
abundance. For magnetic heating, we assume that the magnetic field never rises
above equipartition.  If compression and flux conservation would otherwise
dictate a magnetic field larger than the equipartition value, the field lines
are assumed to reconnect rapidly, converting the magnetic field energy into thermal
energy, thereby re-establishing equipartition conditions.
The cooling function includes a multiple-Gaussian fit to the relevant
cooling emissivities provided by N. Gehrels (see Gehrels \& Williams, 1993, 
and references cited
therein), though with the thermal bremsstrahlung portion supplanted
with more accurate expressions that are valid over a broader range of
physical conditions and with the inclusion of magnetic bremsstrahlung.
For details on the cooling expressions used in the hydrodynamics code
as well as the spectrum calculations below, see Melia \& Coker (1999).
Note that cooling due to Comptonization and any pair production
have not yet been included since they are not thought to be significant in the vicinity
of Sgr A*.  Also, it is assumed that the optical depth is small throughout the volume
of solution.

\subsection{Results of the 3D Simulation}

\begin{figure}[thb]
\epsscale{0.4}
\centerline{\psfig{figure=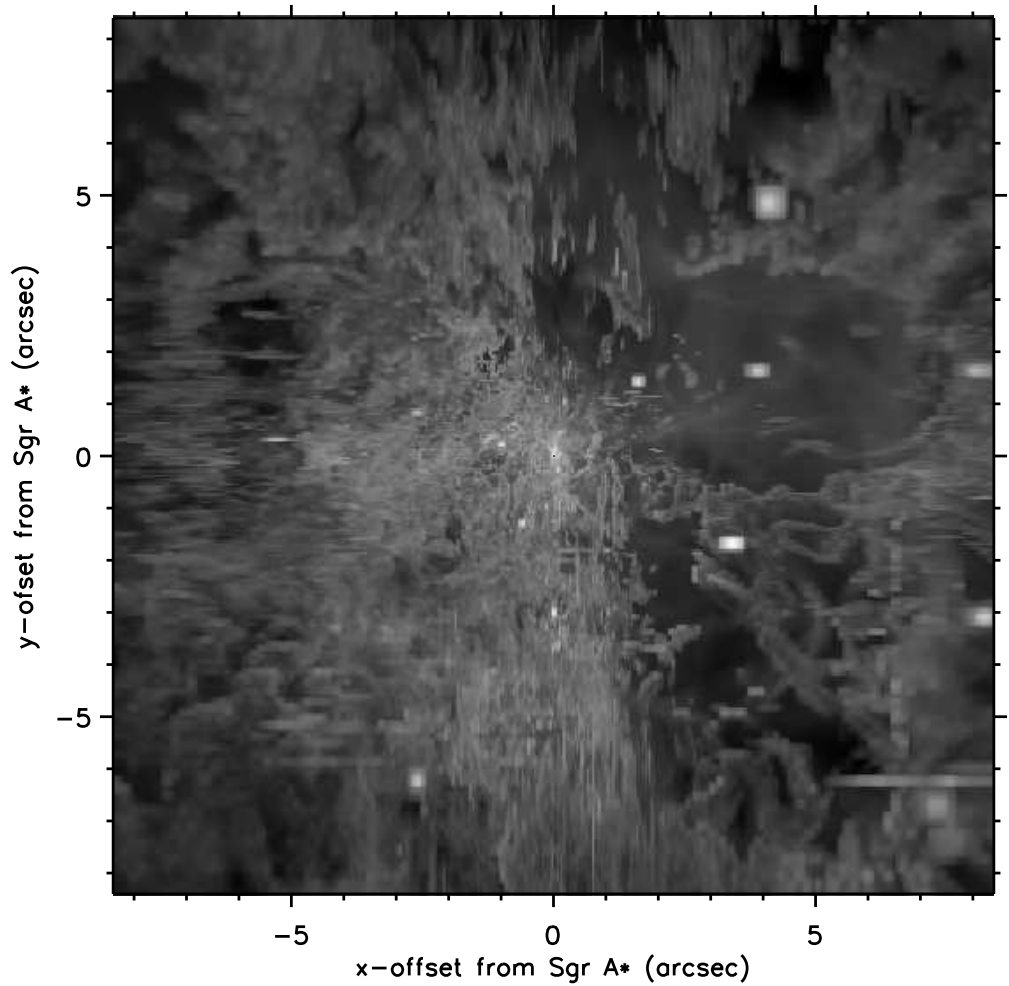,width=\textwidth}}
\caption{The line-of-sight integrated emissivity along the $z$-axis.}
\end{figure}
The integrated emissivity along the line-of-sight,
at $\sim$ 1450 years after the start of the simulation, is shown in Figure 1.
The grey scale is logarithmic
with solid white corresponding to a frequency-integrated intensity of
$\sim1.1\times10^{5}$ erg cm$^{-2}$ s$^{-1}$ steradian$^{-1}$,
and black corresponding to $\sim1\times10^{-2}$ erg cm$^{-2}$ s$^{-1}$ steradian$^{-1}$.
Sgr A* is located at the
center of the image.
Of particular
interest in this image is the appearance of streaks of high-velocity, high-density
gas (``streamers'') that are very reminiscent of features, such as the
so-called ``Bullet'' seen near the Galactic center (Yusef-Zadeh, et al. 1996).
In our simulation, these structures are produced predominantly within the
wind-wind collision regions, and in the future, we shall consider in
greater detail the possibility that the observed high-velocity gas
components near Sgr A* are produced in this fashion. Note
the dominant role played by IRS 13E1 to the lower right of the compact
radio source.

After reaching equilibrium several sound crossing
times ($\sim 1000$ years) after the start of the simulation, the
enclosed mass and energy begin
to fluctuate aperiodically on time scales of less than a few decades
and with an amplitude of up to $50\%$, 
reflecting the turbulent cell nature of
the flow in and out of the central region.
Typically $2.7\times 10^{-3} \;M_\odot$ of gas is trapped within the
cluster at any given time.  Also, although the gas is generally
supersonic, with most of the energy in kinetic form,
the thermal energy can be boosted rather suddenly when the enclosed
magnetic field energy is dissipated.  This occurs when strong shocks
pass through the central region; the shocks compress the field sufficiently
to the point where it reaches, or even surpasses, equipartition and
dissipation ensues. 

\section{The Spectrum of Sgr A*}

\subsection{The Spectrum due to a Dark Cluster}

In order to calculate the observed continuum spectrum, we assume that the observer is
positioned along the negative $z$-axis at infinity and we sum the emission from all zones that
are located at a projected distance, $R_{xy}$, of less than $0.1\,R_A$.
Since the size of Sgr A* (at $\lambda$7 mm) is $\approx 10^{12-13}$ cm (Bower \& Backer, 1998)
and the smallest cell size in our simulation is $7\times10^{14}$ cm, in order
to minimize the inaccuracy due to numerical fluctuations, we have calculated the spectrum
from a central region roughly $10$ times this size ($0.1 R_A$), to include
at least $100$ zones.   Thus clearly our predicted spectrum constitutes
an upper limit to the actual emission expected from Sgr A*.  At the
temperature and density that we encounter here, the dominant components of the continuum
emissivity are electron-ion and electron-electron 
bremsstrahlung; we ignore line emission in these spectral calculations.  

As discussed in Melia (1998), the density in the central region reaches 
a peak value of roughly $10^8$ cm$^{-3}$ and the temperature is never greater
than about $10^8$ K.  Thus, since electrons begin to emit significant
synchrotron radiation only above a few times $10^9$ K, the gas can only
emit cyclotron radiation.  However, the cyclotron emissivity is insignificant
compared to bremsstrahlung so that the final spectrum, as shown in Figure 2,
is a bremsstrahlung spectrum that,in the radio, falls more than 
4 orders of magnitude short of
the observed luminosity of Sgr A* (compare with Figure 6b in Melia, 1998).
Figure 2 shows spectra for 3 points in time during the 3D hydrodynamical simulation.  
The solid, dotted, and dashed curves
are at approximately 1200, 1300, and 1400 years, respectively, after the start of the
simulation.  
The flattening of the density,
temperature, and magnetic field profiles is a direct consequence of the
shallowness of the dark cluster potential compared to the steep potential
gradient encountered by gas falling into a black hole.  The high-energy
shoulder of the emission shown in Figure 3 is characteristic of the highest
temperature $T_{max}$ ($\sim 10^8$ K) attained by the gas and is consistent
with the brightness temperature limit associated with Sgr A*.  Similarly,
the X-ray and $\gamma$-ray emission is significantly below the observed 
upper limits (Predehl \& Truemper, 1994; Goldwurm, et. al., 1994).
\begin{figure}[thb]
\epsscale{0.4}
\centerline{\psfig{figure=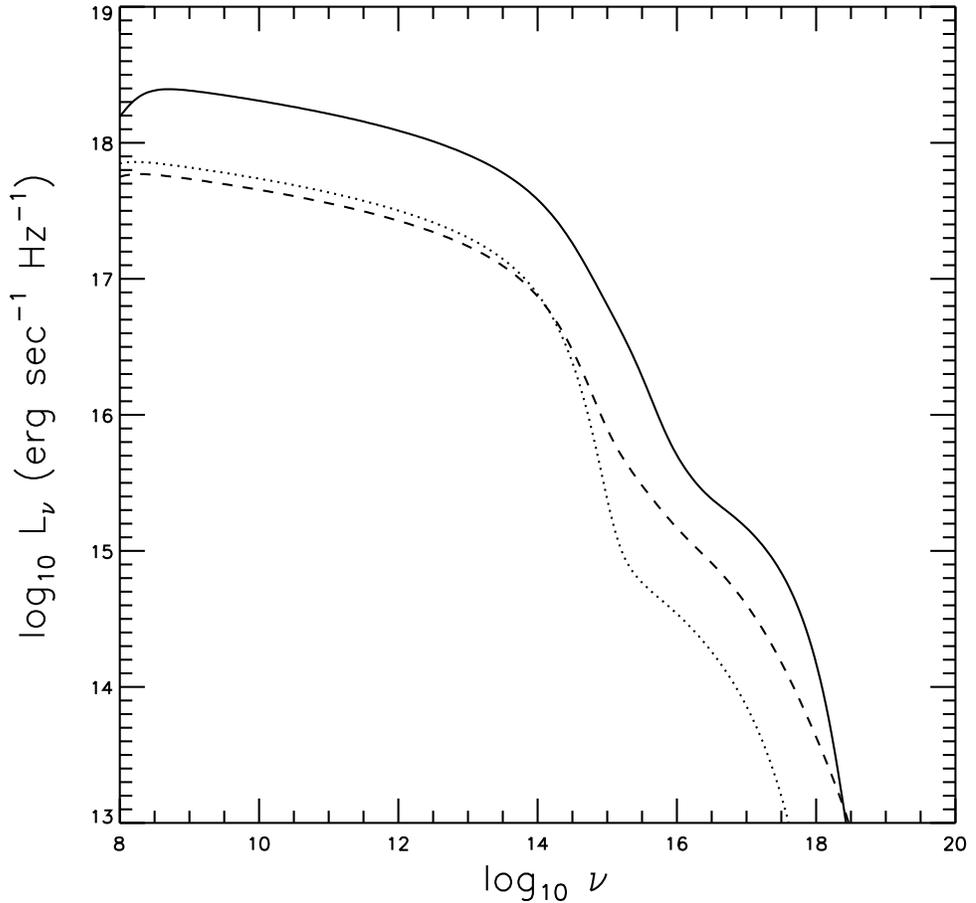,width=\textwidth}}
\caption{Plot of the predicted luminosity density versus frequency.}
\end{figure}

\subsection{The Spectrum due to a Black Hole}

Although we have not yet run a 3D hydrodynamical simulation using the
gravitational potential due to a point mass, we have undertaken 
semi-analytical calculations that improve upon earlier work (Melia, 1994).
Although details will be presented elsewhere (Coker \& Melia, 1999), we
here present some preliminary results.  We now
explicitly solve for the velocity of the spherical flow using the relativistic
Euler equation and integrate over $\mu$, the cosine of the angle between
the line-of-sight and the flow.  We also include effects due to inverse Compton
scattering and the index of refraction.

In recent work
(Lo, et. al., 1998) it has been suggested that, at mm wavelengths at least, the intrinsic
size of Sgr A* varies as
$\nu^\alpha$ with $-1.9 < \alpha < -0.7$.  It is difficult to
compare the absolute observed size, which is usually the FWHM of a best-fit Gaussian
(e.g. Krichbaum, et. al., 1998), to a theoretical size, which is usually
given as the surface of last scatterring where $\tau(\nu) \equiv 1$; the observational
size will tend to be larger than the theoretical size with these definitions.  
However, Figure 3 shows a plot of the predicited {\sl intrinsic} size of Sgr A*
along with present observational values (see Lo, et. al. 1998 and references
cited therein for details).  The slope of the line is $\sim -0.6$,
somewhat less steep than the observed value.  This is not surprising since this particular
model does not produce enough low frequency magnetic bremsstrahlung at large
radii.  The slope should steepen when a more optimal fit to the spectrum is found.  
Note that the lower size bound on the plot corresponds to the
size of the black hole itself ($1 R_s \sim 6 \mu$as).  
\begin{figure}[thb]
\epsscale{0.4}
\centerline{\psfig{figure=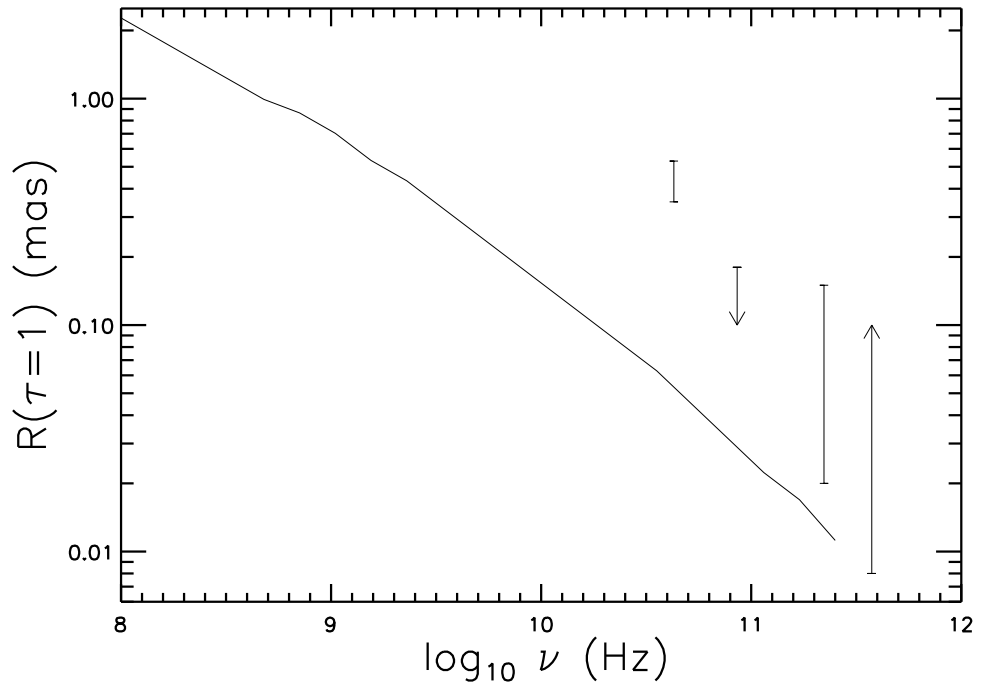,width=\textwidth}}
\caption{Plot of the predicted size of Sgr A* versus frequency (Coker \& Melia, 1999).}
\end{figure}

\section{Summary}
It does not appear, based on these calculations, that the gravitational 
potential of a distributed dark mass, be it due to a compact cluster
of stellar remnants or an even more exotic collection of objects (e.g.
Tsiklauri \& Viollier, 1998), can compress the gas from stellar winds to
the point where the temperature, density, and magnetic field can produce
an observationally significant cyclotron/synchrotron emissivity at GHz
frequencies.  Although the spatial resolution of the simulation near the
origin can be improved over that used here, it is unlikely that it will
alter the results; the lack of compression is due to the inherently flat
potential of any distributed dark cluster.  Thus, aside from the issue
of whether any dark cluster can account for the observed GC potential and
whether any such cluster is stable over a significant fraction of the age
of the Galaxy, our 3D hydrodynamical simulation has shown that the radio
emissivity from the gas trapped within any such cluster cannot reprodude
the spectrum of Sgr A*.  We are left to conclude that either Sgr A* is
unrelated to the accreting or trapped GC gas or that Sgr A* is the signature
of an accreting, massive black hole.  However, models of the structure and
mechanism of radio emission from such a black hole are now faced with
futher contraints due to recent observations of the intrinsic source 
structure of Sgr A*.  Coupled with the observed spectral energy
distribution of Sgr A*, we may be close to differentiating between the
various accretion models.

\acknowledgments
This work was partially supported by NASA grant NGT-51637
and has made use of NASA's Astrophysics Data System Abstract Service.

\end{document}